# Low electrical conductivity of dry $CaSiO_3$ perovskite under lower mantle conditions

Yoshiyuki Okuda[1,2,*], Bin Chen[1], Juliana Peckenpaugh[1,3], Hirokazu Kadobayashi[4]

[1] *Hawai'i Institute of Geophysics and Planetology, University of Hawai'i at Manoa, Honolulu, Hawai'i 96822, USA*

[2] *Department of Earth and Planetary Sciences, Institute of Science Tokyo, Meguro, Tokyo 152-8551, Japan*

[3] *Department of Earth Sciences, University of Hawai'i at Manoa, Honolulu, Hawai'i 96822, USA*

[4] *SPring-8, Japan Synchrotron Radiation Research Institute, Sayo, Hyogo 679-5198, Japan*

*Corresponding author

*E-mail:* yokuda@hawaii.edu (Y. Okuda)

## ABSTRACT

Electrical conductivity (EC) provides important constraints on the composition and volatile distribution of Earth's deep mantle, yet the EC of davemaoite ($CaSiO_3$ perovskite), a major lower-mantle phase, remains poorly constrained. We measured the EC of nominally dry $CaSiO_3$ perovskite at pressures up to 89 GPa and temperatures up to 2200 K using impedance spectroscopy in a laser-heated diamond anvil cell. Conductivity increases with temperature but decreases systematically with pressure and it is substantially lower than previously reported, yet broadly consistent with recent theoretical predictions for oxygen-vacancy-mediated ionic transport. A distinct change in the temperature dependence coincides with the tetragonal-to-cubic phase boundary, revealing a modest enhancement of ionic transport across the structural transition. Along a normal lower-mantle geotherm, dry davemaoite is comparable in conductivity to bridgmanite near the top of the lower mantle but becomes progressively less conductive with depth. Under cold-slab conditions, dry davemaoite is substantially less conductive than dry subducted MORB and cannot account for the observed high-conductivity anomalies. Dry davemaoite therefore contributes little to bulk lower-mantle conductivity.

## Highlights

- We measured the electrical conductivity of $CaSiO_3$ perovskite up to 89 GPa and 2200 K.
- Conductivity decreases with pressure, consistent with oxygen-vacancy transport.
- Dry davemaoite is substantially less conductive than previously reported.
- Conductivity change concurs with the tetragonal-to-cubic phase transition.
- Dry davemaoite cannot explain conductivity anomalies beneath subduction zones.

## 1. Introduction

Electrical conductivity (EC) is a geophysical observable highly sensitive to the composition and volatile content of Earth's deep interior. Electromagnetic sounding studies have revealed localized high-conductivity anomalies in the lower mantle, particularly beneath subduction zones (Kelbert et al., 2009; Kuvshinov, 2012; Shimizu et al., 2010; Zhang and Yang, 2022). Identifying the mineralogical origin of these anomalies is essential for understanding deep volatile cycling, and several candidate phases have been proposed. Dry, Fe-free $MgSiO_3$ melt exhibits relatively low conductivity (Okuda et al., 2024), whereas hydrous silicate melts can be substantially more conductive and may account for mantle high-conductivity anomalies even at small melt fractions (Yoshino and Xie, 2026). Carbonate melts have also been suggested as highly conductive interconnected phases capable of producing localized EC anomalies (Zhao et al., 2026). Hydrous lower-mantle minerals are another promising candidates. Recent high-pressure experiments have shown that hydrous bridgmanite can exhibit EC up to an order of magnitude higher than dry bridgmanite and other nominally dry lower-mantle assemblages (Hiraoka et al., 2026; Ohta et al., 2017, 2008; Trubowitz et al., 2026), owing to efficient proton conduction (Okuda et al., 2026b). Hydrous $SiO_2$ may also become superionic under lower-mantle conditions, increasing its conductivity by approximately two orders of magnitude and potentially accounting for high-conductivity anomalies (Inada et al., 2025).

Davemaoite ($CaSiO_3$ perovskite) is another potentially important contributor to lower mantle EC. Although generally regarded as the third most abundant mineral in a pyrolitic lower mantle, it may be depleted in the ambient lower mantle (Hao et al., 2026); it is, however, a major constituent of subducted basaltic oceanic crust under lower-mantle conditions (Hirose et al., 2005). Its geophysical importance has motivated extensive investigations of its physical properties, including seismic velocities (Gréaux et al., 2019; Thomson et al., 2019; Zhou et al., 2025) and thermal conductivity (Zhang et al., 2021). Previous experimental and theoretical studies suggest that davemaoite undergoes a ferroelastic tetragonal-to-cubic phase transition at elevated temperatures, although the transition temperature remains debated (Adams and Oganov, 2006; Komabayashi et al., 2007; Kurashina et al., 2004; Okuda et al., 2026a; Sagatova et al., 2021; Stixrude et al., 2007; Wu et al., 2024; Zhang et al., 2025). Constraining its transport properties across this transition is therefore important for understanding the geophysical behavior of davemaoite under lower-mantle conditions.

Despite its importance, the EC of davemaoite remains poorly constrained. To date, only one experimental study has measured its EC under simultaneous high-pressure and high-temperature (*P*–*T*) conditions (Fei et al., 2017). That study reported that davemaoite has the highest EC among the major lower-mantle minerals and proposed that it could contribute substantially to local high-conductivity anomalies in the topmost lower mantle. This interpretation contrasts, however, with first-principles calculations that predict substantially lower oxygen-ion conductivity (Adams and Oganov, 2006; Schulze and Steinle-Neumann, 2026). This discrepancy remains unresolved, largely because no independent high-*P*–*T* EC measurements of davemaoite exist.

Here, we report *in situ* EC measurements of davemaoite to 89 GPa and 2200 K using impedance spectroscopy in a laser-heated diamond anvil cell (LHDAC). We find that dry davemaoite is substantially less conductive than previously reported, in agreement with recent theoretical predictions. Dry davemaoite alone therefore cannot account for the observed lower-mantle high-conductivity anomalies, which instead require additional conductive components such as carbonate melts or hydrous phases (Inada et al., 2025; Zhao et al., 2026). We further observe a distinct change in the temperature dependence of conductivity that coincides with the reported tetragonal-to-cubic phase boundary, suggesting a link between crystal symmetry and oxygen-ion transport in davemaoite.

## 2. Experimental Methods

### *2.1. Cell configuration for high P-T impedance measurements*

Symmetric LHDACs with 200 μm culet diamond anvils were used in all experiments. A hole matching the culet diameter was drilled at the center of a rhenium gasket pre-indented to ~30 μm. A mixture of cubic boron nitride (cBN) + $TiO_2$ (binder) was loaded into the hole and subsequently compressed to ~45 μm in thickness (Okuda et al., 2022). After compaction, a hole with ~1/3 the diameter of the culet was drilled in the cBN mixture using a UV laser to form the sample chamber. We used nominally dry $CaSiO_3$ glass powder (Alfa Aesar, 89741) as a starting material for $CaSiO_3$ perovskite. $CaSiO_3$ glass was formed into a pellet and sandwiched between (1) iridium-foil electrodes (99.8% purity, Nilaco Corp.; approximately 1–2 μm thick and 50 μm wide), (2) a 1.6 μm-thick boron-doped diamond (BDD) layer used as a laser absorber in run #1, which was omitted in run #2, and (3) 10 μm-thick $Al_2O_3$ thermal insulation layers (Fig. 1). The iridium foils were connected to platinum foil leads on both sides of the gasket, which in turn were connected to copper lead wires and the impedance analyzer. Pressure was determined using the thermal equation of state (EoS) of $CaSiO_3$ (Okuda et al., 2026a). The resulting pressures agreed with those determined using the thermal EoS of iridium (Anzellini et al., 2021). No contamination by Ir or other materials was observed in the recovered samples (Fig. S1).

All experiments were performed at SPring-8, BL10XU (Hirao et al., 2020). In each run, the starting material was first compressed in an LHDAC to the initial target pressure within the stability field of davemaoite and then heated to approximately 2000 K and held for 15 min to synthesize davemaoite. Formation of davemaoite was confirmed by *in situ* XRD measurements (Fig. 2). A monochromatic ~30 keV beam was used, and the patterns were fitted using PDIndexer (Seto et al., 2010). The electrical impedance was measured using HIOKI 3532-80 with a Four-Terminal Probe 9500 (Hioki Corp.). Two 100 W single-mode Yb fiber lasers (YLR-100, IPG Photonics) were used to heat the sample in a DAC from both sides. At lower laser powers, where thermal emission was insufficient for direct pyrometric measurement, temperatures were estimated from linear least-squares relationships between applied laser power and measured temperature established at higher powers (Fig. S2). We employed beam-shaping optics to produce a flat energy distribution. Temperature at the surface of the electrode was determined by a spectro-radiometric method. Temperature distributions in the sample assembly were calculated using a two-dimensional steady-state heat-conduction model implemented in Python for the five pressure conditions listed in Table 1. Laser heating was represented by a uniform flat-top surface heat flux applied over a lateral width equal to the measured laser-spot diameter on both the upper and lower Ir surfaces. For each run, the heat flux was adjusted so that the maximum Ir-surface temperature matched the measured peak temperature. The model included heat conduction through the $CaSiO_3$ perovskite sample, 1 μm-thick Ir layers, and the surrounding pressure medium, with the outer boundaries of the computational domain fixed at 300 K. The laser-spot diameter, peak surface temperature, and sample thickness were set to the measured values for each run (Table 1). The thermal conductivity of $CaSiO_3$ was estimated from the pressure-dependent relationship reported by Zhang et al. (2021), including its $T^{-1.11}$ temperature dependence. Thermal conductivities of alumina and Ir were estimated from Hofmeister (1999) and Bhatt et al. (2024), respectively, with the Ir conductivity extrapolated to our highest investigated pressure of ~89 GPa from the reported pressure dependence up to 60 GPa. The heat-conduction equation was solved on a uniform Cartesian finite-difference grid with a spatial resolution of 0.1 μm. The representative sample temperature was defined as the two-dimensional area average within the laser-heated region, with an estimated uncertainty of approximately ±10% (see Fig. S3 for the simulation results).

For impedance measurements, the alternating voltage was set to 0.5 V, and the frequency range to $10^1$–$10^6$ Hz. The impedance spectra, represented as Nyquist plots, exhibited a semicircular arc corresponding to the bulk response, followed by an inclined low-frequency tail attributed to electrode polarization associated with ion migration (Fig. 3; see Results for details). The spectra were fitted using RelaxIS-3 software (rhd instruments GmbH & Co. KG) with the following equivalent circuit:

$$C_{\mathrm{hf}} \parallel (R_{\mathrm{b}} - \mathrm{CPE}_{\mathrm{int}}) \parallel \mathrm{CPE}_{\mathrm{b}}, \quad (2)$$

where the hyphen denotes elements connected in series and ∥ denotes parallel branches, $C_{\mathrm{hf}}$ represents the high-frequency capacitive response, $\mathrm{CPE}_{\mathrm{b}}$ describes the distributed dielectric relaxation of the sample, $R_{\mathrm{b}}$ is the bulk resistance, and $\mathrm{CPE}_{\mathrm{int}}$ accounts for nonideal low-frequency interfacial polarization caused by ionic accumulation at the nominally blocking electrode–sample interfaces. The equivalent-circuit model reproduced the measured spectra well over the investigated frequency range (Fig. 3).

The determination of EC ($\sigma$) from measured electrical resistance requires the sample thickness ($L$) and the surface area ($S$) as:

$$\sigma = \frac{L}{RS}\,. \quad (3)$$

The recovered sample thickness was measured on a cross section prepared using a focused Ga ion beam (FIB, FEI, Versa 3D DualBeam). To estimate the sample thickness at the maximum pressure of each run, the recovered thickness was corrected for elastic decompression using the EoS of $CaSiO_3$ perovskite (Okuda et al., 2026a). The pressure dependence of the sample thickness was estimated from micrometer measurements of the total sample-chamber thickness at several pressures, assuming negligible diamond deformation and uniform compression (Sinmyo et al., 2014). The resulting pressure-dependent thickness estimates were generally consistent with the trend reported by Sinmyo et al. (2014) (Fig. S4). Regarding the surface area, we assumed that the electric current induced by an applied electric field flows only in the laser-heated portion. The $S$ value at high temperatures was therefore considered identical to the laser spot size. The overall uncertainty in EC, $u_{EC}$, was calculated as:

$$u_{EC} = \sqrt{u_L^2 + u_S^2 + u_R^2} \quad (4)$$

in which $u_L$ is the standard deviation of the observed sample thickness (~20%), $u_S$ is the uncertainty in the surface area (~10%), and $u_R$ is the fitting uncertainty (a few %). The resulting relative uncertainty in EC was typically ~25–40% (Table 1).

## 3. Results

The impedance spectra consisted of a semicircle attributed to the bulk response of $CaSiO_3$ perovskite, followed by an inclined low-frequency tail that deviated from the 45° slope characteristic of Warburg diffusion (Fig. 3). Similar low-frequency features were previously reported for $CaSiO_3$ perovskite and interpreted as electrode polarization caused by the accumulation of mobile ionic defects at blocking electrodes (Fei et al., 2017). Because oxygen-vacancy migration has been proposed as the dominant conduction mechanism in $CaSiO_3$ perovskite (Adams and Oganov, 2006; Schulze and Steinle-Neumann, 2026; Wang and He, 2025), the observed low-frequency response is consistent with electrode polarization at the Ir electrodes by the accumulation of oxygen vacancies. With increasing temperature, the bulk semicircle shifted progressively beyond the measured frequency window because the characteristic relaxation frequency, $f = 1/(2\pi RC)$, increases as resistance decreases.

Figure 4 shows the high *P-T* EC of $CaSiO_3$ perovskite. In run #1, EC was measured at pressures of approximately 46, 53, and 86 GPa. At all pressures, conductivity increased with increasing temperature. At approximately 46 GPa, conductivity increased from $9.8\times10^{-4}$ S/m at 1040 K to $2.1\times10^{-1}$ S/m at 1890 K. The data followed an approximately linear Arrhenius-type trend between 1040 and 1440 K (Region I), showed a stronger temperature dependence between 1440 and 1680 K (Region II), and returned to a weaker temperature dependence above 1680 K (Region III). At approximately 53 GPa, conductivity increased from $4.4\times10^{-4}$ S/m at 1080 K to $1.2\times10^{-1}$ S/m at 2010 K and was lower than that measured at approximately 46 GPa at comparable temperatures. A similar three-stage temperature dependence was observed, with an approximately linear Arrhenius-type trend between 1080 and 1550 K, a stronger temperature dependence between 1550 and 1770 K, and a weaker temperature dependence above 1770 K. The lowest conductivities were obtained at approximately 86 GPa, where a similar temperature-dependent behavior was observed. At comparable temperatures, conductivity decreased systematically with increasing pressure. This negative pressure dependence was reproduced in a separate experimental run at approximately 31 and 67 GPa (run #2). At 31 GPa, the data followed an approximately linear Arrhenius-type trend over the measured temperature range. In contrast, the data at 67 GPa showed three distinct temperature regimes, consisting of an initial Arrhenius-type trend, an intermediate interval with stronger temperature dependence, and a return to a weaker temperature dependence at higher temperatures, similar to the behavior observed in run #1.

Our measurements are broadly consistent with the conductivity trends calculated by Schulze and Steinle-Neumann (2026). Most Region I data overlap the calculated values at comparable pressures within experimental uncertainty, although the 86 GPa data are slightly more conductive than the corresponding calculation. In contrast, the Region III conductivities lie consistently above the calculated trends. At 31 GPa, the measured conductivity is approximately a factor of 4–6 lower than the previous experimental values reported at 17 and 24 GPa by Fei et al. (2017) over the overlapping temperature range.

## 4. Discussion

### *4.1. Temperature-dependent electrical conductivity and the tetragonal-to-cubic transition*

The temperature dependence within each regime was fitted using the Arrhenius relation:

$$\sigma = \sigma_0 \exp\left(-\frac{\Delta H}{\mathrm{k}T}\right), \tag{5}$$

where $\sigma_0$ is the pre-exponential factor, $\Delta H$ is the activation enthalpy, and k is the Boltzmann constant. Linear least-squares fits were performed separately for Regions I, II, and III. Region I yielded $\Delta H$ values of 1.23 ± 0.03, 0.95 ± 0.02, 0.95 ± 0.02, 1.14 ± 0.45, and 1.15 ± 0.04 eV at 31, 46, 53, 67, and 86 GPa, respectively. Because the Region I fit at 67 GPa contained only two data points, its uncertainty was estimated by propagating the uncertainties in log $\sigma$. In contrast, Region II exhibited substantially larger $\Delta H$ values of 1.45 ± 0.19, 2.05 ± 0.19, 4.07 ± 0.22, and 2.32 ± 0.65 eV at 46, 53, 67, and 86 GPa, respectively. At higher temperatures, Region III returned to lower $\Delta H$ values of 1.04 ± 0.15, 1.05 ± 0.10, and 1.60 ± 0.04 eV at 46, 53, and 67 GPa, respectively (Table 2). The transition temperatures between Regions I and II closely follow the tetragonal-to-cubic phase boundary of $CaSiO_3$ perovskite determined by high-$P$–$T$ X-ray diffraction measurements (Okuda et al., 2026a) and are broadly consistent with previous experimental and theoretical constraints (Adams and Oganov, 2006; Stixrude et al., 2007, 1996; Thomson et al., 2019; Zhang et al., 2025) (Fig. 5). This suggests that Region I represents conduction in tetragonal $CaSiO_3$ perovskite, whereas Region III represents conduction in the high-temperature cubic phase. Region II, characterized by anomalously large apparent activation enthalpies, likely reflects behavior associated with the tetragonal-to-cubic transformation. The anomalous behavior may reflect either intrinsic transport processes associated with the phase transition or apparent broadening caused by temperature gradients within the sample. The cubic regime is systematically more conductive than the extrapolation of the low-temperature tetragonal Arrhenius trend, indicating a modest enhancement in ionic transport across the tetragonal-to-cubic transition. The $\Delta H$ values of Regions I and III overlap within uncertainty at 46, 53, and 67 GPa (Fig. S5), suggesting that the transition does not substantially modify the activation barrier for conduction. The relatively small changes in both conductivity and $\Delta H$ may reflect the weak structural contrast between the two phases at high temperature, as the tetragonal distortion diminishes progressively toward the cubic phase boundary (Zhang et al., 2025). The $\Delta H$ values in Regions I and III are comparable to previous experimental and theoretical estimates for oxygen-ion transport in davemaoite (Fig. S5), consistent with oxygen-vacancy-mediated ionic transport in both the tetragonal and cubic regimes.

Our measured conductivity decreases systematically with increasing pressure, consistent with the calculated behavior of extrinsic oxygen-vacancy diffusion in davemaoite (Schulze and Steinle-Neumann, 2026). This contrasts with the slightly positive pressure dependence reported by Fei et al. (2017), which corresponds to a negative activation volume, atypical of ionic conduction. Notably, although their 17 GPa conductivities are broadly comparable in magnitude to the calculated values (Schulze and Steinle-Neumann, 2026), the 24 GPa data are substantially higher and largely account for the opposite pressure dependence (Fig. 4). Fei et al. (2017) also noted that their samples may have contained structural OH, which could have affected the conductivity. The discrepancy may therefore reflect sample chemistry, an anomalously high conductivity in the 24 GPa dataset, or both, rather than an intrinsic positive pressure dependence of oxygen-vacancy conduction.

*4.2. The electrical conductivity of dry $CaSiO_3$ perovskite in the ambient lower mantle*

To evaluate the geophysical significance of dry davemaoite, we calculated its EC along the normal lower-mantle geotherm of Katsura (2022) using the Arrhenius relations obtained in this study. At approximately 31 GPa, where only a single Arrhenius trend was observed in the cubic regime, the measured relation was used because davemaoite is expected to be cubic under normal lower-mantle conditions (Okuda et al., 2026a). Along the geotherm, the ECs at 31, 46, 53, and 67 GPa are 0.84, 0.43, 0.19, and 0.63 S/m, respectively (Fig. 6). The 31 GPa value is comparable to the globally averaged mantle EC inferred from geomagnetic observations (Kuvshinov et al., 2021; Velímský and Knopp, 2021), as well as to the ECs of other major lower-mantle minerals, including (Fe,Al)-bearing bridgmanite (Sinmyo et al., 2014; Xu and McCammon, 2002) and ferropericlase (Hiraoka et al., 2026), under comparable *P–T* conditions. The values at higher pressures, by contrast, fall systematically below the geophysically inferred mantle conductivity and generally below those reported for other major lower-mantle minerals and assemblages (Fig. 6). At approximately 86 GPa, extrapolation of the Region II Arrhenius relation to the corresponding geotherm temperature of 2399 K yields an EC of approximately 0.45 S/m. However, the Region III relation was not experimentally constrained at this pressure, and the geotherm temperature lies above the maximum experimental temperature. In our datasets at 46–67 GPa, Region III exhibits a weaker temperature dependence than Region II; therefore, if similar behavior occurs at ~86 GPa, extrapolation of the steeper Region II relation to 2399 K would likely overestimate the actual conductivity. Nevertheless, even this potentially overestimated value remains lower than the geophysically inferred mantle conductivity. Because davemaoite is comparable in conductivity to bridgmanite near 31 GPa but becomes substantially less conductive at higher pressures, its effect on bulk conductivity remains negligible even at a nominal pyrolitic abundance of 7 vol% (Fig. S6). This contribution would be further reduced if davemaoite is depleted in the ambient lower mantle relative to its nominal pyrolitic abundance (Hao et al., 2026). Thus, regardless of the topology of davemaoite, the EC of the pyrolitic lower mantle remains essentially controlled by bridgmanite, except where ferropericlase forms a locally interconnected network (Hiraoka et al., 2026).

*4.3. The electrical conductivity of dry $CaSiO_3$ perovskite in subducted oceanic crust*

Davemaoite is substantially more abundant in subducted basaltic oceanic crust, accounting for approximately 23 wt% of a MORB assemblage (Hirose et al., 2005). We therefore estimated its EC along the cold slab geotherm (Ohtani, 2020). The calculated ECs of dry davemaoite at representative slab temperatures are approximately $8.3 \times 10^{-4}$, $6.8 \times 10^{-3}$, $3.5 \times 10^{-3}$, $5.3 \times 10^{-3}$, and $1.5 \times 10^{-2}$ S/m at 31, 46, 53, 67, and 86 GPa, respectively (Fig. 7). At 86 GPa, the representative slab temperature is approximately 1840 K, which lies within the experimentally constrained Region II temperature range of 1800–1880 K. Therefore, the calculated EC at this pressure does not require extrapolation of the Arrhenius relation. These conductivities are substantially lower than the approximately 0.5–5 S/m estimated for dry subducted MORB over comparable lower-mantle depths (Inada et al., 2025; Ohta et al., 2010) and are also far below the high-conductivity anomalies observed beneath northeastern China and the Japan Sea (Kelbert et al., 2009; Shimizu et al., 2010). Thus, despite its relatively high abundance in subducted basaltic crust, dry davemaoite does not substantially enhance the bulk EC of subducted oceanic crust and is unlikely to account for the observed high-EC anomalies.

Our results provide an end-member constraint for dry, compositionally pure $CaSiO_3$ davemaoite. Natural davemaoite in subducted oceanic crust, however, can incorporate appreciable amounts of Al and Ti, and the effects of these substitutions on its EC remain experimentally unconstrained. Davemaoite in MORB compositions is expected to contain up to approximately 4–5 wt% $Al_2O_3$ (Hirose et al., 2005; Kurashina et al., 2004). Aliovalent substitution of $Al^{3+}$ for $Si^{4+}$ has been shown to involve oxygen-vacancy formation in Al-bearing davemaoite (Bläß et al., 2007). Such a substitution mechanism would increase the oxygen-vacancy concentration and thereby the density of ionic charge carriers, which is expected to enhance EC according to the Nernst–Einstein relation. Al incorporation also strongly affects the structural transition. At approximately 50 GPa, addition of 5.9 wt% $Al_2O_3$ changes the low-temperature structure from tetragonal to orthorhombic and raises the transition temperature to the cubic phase by 1260 K compared to pure $CaSiO_3$ (Kurashina et al., 2004). Ti can also be incorporated into davemaoite in subducted MORB, with approximately 1.2–1.3 wt% $TiO_2$ reported (Ricolleau et al., 2010). Ti incorporation stabilizes the tetragonal structure (Chao et al., 2024) and substantially modifies the tetragonal-to-cubic transition behavior of davemaoite (Okuda et al., 2026a). Because our results indicate enhanced EC across the structural transition, Al- and Ti-induced shifts of the transition boundary could modify davemaoite conductivity in subducted oceanic crust. Direct EC measurements of Al- and Ti-bearing davemaoite are therefore needed to evaluate its contribution to the bulk EC of subducted slabs. Hydrogen may also strongly influence the EC of davemaoite, as even trace $H_2O$ can substantially enhance electrical transport in nominally anhydrous minerals. The water-storage capacity of davemaoite remains uncertain, with reported estimates ranging from approximately 100–1000 ppm $H_2O$ (Hirschmann, 2006; Keppler and Bolfan-Casanova, 2006; Németh et al., 2017) to >4000 ppm $H_2O$ (Chen et al., 2020; Murakami et al., 2002). A more recent study constrained the $H_2O$ content of davemaoite to approximately 0.04–0.08 wt% under lower-mantle conditions (Ishii et al., 2026). Even such low $H_2O$ contents may substantially enhance EC: only ~200–300 ppm $H_2O$ markedly increases the EC of bridgmanite (Okuda et al., 2026b). Because subducted oceanic crust in the lower mantle may be rehydrated by fluids released through dehydration of the underlying slab mantle (Walter, 2021), davemaoite in subducted MORB may contain $H_2O$ at comparable levels. Future measurements of hydrous davemaoite will therefore be important for quantitatively evaluating subduction-related lower-mantle conductivity anomalies and the role of subducted oceanic crust in Earth's deep water cycle.

## 5. Conclusions

We determined the EC of dry, end-member davemaoite at lower-mantle pressures using impedance spectroscopy in an LHDAC. The measured conductivities are substantially lower than previously reported and decrease systematically with increasing pressure, in broad agreement with theoretical predictions for oxygen-vacancy-mediated ionic transport. A distinct change in the temperature dependence of conductivity coincides with the tetragonal-to-cubic phase boundary, indicating a modest enhancement of ionic transport across the structural transition. Along a normal lower-mantle geotherm, dry davemaoite is comparable in conductivity to bridgmanite near the top of the lower mantle but becomes progressively less conductive with depth, limiting its contribution to bulk mantle EC. Under cold-slab conditions, its conductivity is substantially lower than that inferred for subducted MORB and cannot account for the high-conductivity anomalies observed beneath subduction zones. These results establish dry $CaSiO_3$ davemaoite as an end-member constraint for evaluating the effects of Al, Ti, and $H_2O$ on the conductivity of compositionally realistic davemaoite in the lower mantle.

**CRediT authorship contribution statement**

**Yoshiyuki Okuda:** Conceptualization, Methodology, Investigation, Writing – original draft, Writing – review & editing, Funding acquisition. **Bin Chen:** Writing – review & editing, Funding acquisition. **Juliana Peckenpaugh:** Investigation, Writing – review & editing. **Hirokazu Kadobayashi:** Resources, Writing – review & editing.

## Declaration of competing interest

The authors declare that they have no competing interests.

## Data and materials availability

All data are available in the main text or the supplementary materials.

## Acknowledgments

This work was supported by the JSPS KAKENHI Grant No. 22J00928 (Y.O.), 26K17248 (Y.O.), and NSF Grants EAR-2447900, EAR-2127807 (B.C.). The present measurements were carried out at the beamline BL10XU, SPring-8 (proposal no. 2025A1259; no. 2025B1276).

## Supplementary materials

Figs. S1–S6; Supplementary Text S1.

**Table 1. Experimental conditions and electrical conductivity of $CaSiO_3$ perovskite.**

| Run # | Pressure (GPa) | Measured $T$ (K) | Simulated sample $T$ (K) | Sample thickness (μm) | Laser spot size (μm) | log [$R$ (Ohm)] | log [$\sigma$ (S/m)] |
|---|---|---|---|---|---|---|---|
| 1 | 48(5) | 1690(170) | 1560(160) | 20.8(29) | 30.0(30) | 5.85(2) | -1.38(13) |
| | 48(5) | 1950(190) | 1800(180) | | | 5.34(1) | -0.87(13) |
| | 47(5) | 2030(200) | 1890(190) | | | 5.14(2) | -0.67(13) |
| | 46(5) | 1960(200) | 1820(180) | | | 5.25(2) | -0.78(13) |
| | 45(4) | 1820(180) | 1680(170) | | | 5.49(1) | -1.02(13) |
| | 45(4) | 1760(180) | 1620(160) | | | 5.61(1) | -1.14(13) |
| | 45(4) | 1710(170) | 1570(160) | | | 5.78(1) | -1.31(13) |
| | 45(4) | 1630(160) | 1490(150) | | | 6.01(2) | -1.54(13) |
| | 45(4) | 1570(160) | 1440(140) | | | 6.20(2) | -1.73(13) |
| | 45(4) | 1510(150) | 1380(140) | | | 6.34(2) | -1.87(13) |
| | 44(4) | 1440(140) | 1300(130) | | | 6.57(2) | -2.10(13) |
| | 44(4) | 1370(140) | 1240(120) | | | 6.77(4) | -2.30(14) |
| | 44(4) | 1300(130) | 1170(120) | | | 7.00(5) | -2.53(14) |
| | 43(4) | 1230(120) | 1100(110) | | | 7.22(6) | -2.75(15) |
| | 43(4) | 1170(120) | 1040(100) | | | 7.48(6) | -3.01(15) |
| | | | | | | | |
| 1 | 57(6) | 2140(210) | 2010(200) | 14.7(21) | 30.0(30) | 5.25(2) | -0.93(13) |
| | 55(5) | 2030(200) | 1910(190) | | | 5.37(2) | -1.05(13) |
| | 53(5) | 1970(200) | 1850(190) | | | 5.47(3) | -1.16(14) |
| | 52(5) | 1890(190) | 1770(180) | | | 5.61(2) | -1.29(13) |
| | 52(5) | 1840(180) | 1720(170) | | | 5.78(2) | -1.46(14) |
| | 53(5) | 1800(180) | 1680(170) | | | 5.98(2) | -1.66(13) |
| | 53(5) | 1740(170) | 1620(160) | | | 6.14(2) | -1.82(13) |
| | 53(5) | 1670(170) | 1550(160) | | | 6.32(2) | -2.00(13) |
| | 52(5) | 1610(160) | 1490(150) | | | 6.46(3) | -2.15(14) |
| | 52(5) | 1560(160) | 1440(140) | | | 6.60(4) | -2.29(14) |
| | 51(5) | 1490(150) | 1380(140) | | | 6.75(3) | -2.43(14) |
| | 51(5) | 1430(140) | 1310(130) | | | 6.94(4) | -2.62(14) |
| | 51(5) | 1370(140) | 1260(130) | | | 7.08(4) | -2.77(14) |
| | 50(5) | 1320(130) | 1200(120) | | | 7.25(5) | -2.93(14) |
| | 50(5) | 1250(120) | 1140(110) | | | 7.46(4) | -3.14(14) |
| | 50(5) | 1190(120) | 1080(110) | | | 7.68(6) | -3.36(15) |
| | | | | | | | |
| 1 | 89(9) | 1830(180) | 1740(170) | 6.5(9) | 30.0(30) | 6.09(2) | -2.13(13) |
| | 88(9) | 1900(190) | 1800(180) | | | 5.97(2) | -2.01(13) |
| | 88(9) | 1970(200) | 1880(190) | | | 5.67(2) | -1.70(13) |
| | 87(9) | 1910(190) | 1810(180) | | | 5.85(2) | -1.89(13) |
| | 86(9) | 1800(180) | 1700(170) | | | 6.05(3) | -2.08(14) |
| | 85(9) | 1760(180) | 1660(170) | | | 6.23(2) | -2.26(13) |
| | 85(8) | 1650(170) | 1560(160) | | | 6.43(3) | -2.47(14) |
| | 84(8) | 1570(160) | 1480(150) | | | 6.66(3) | -2.70(14) |
| | 84(8) | 1480(150) | 1380(140) | | | 6.96(3) | -3.00(14) |
| | 84(8) | 1400(140) | 1300(130) | | | 7.21(4) | -3.25(14) |
| | 83(8) | 1310(130) | 1210(120) | | | 7.45(5) | -3.49(14) |
| | | | | | | | |
| 2 | 32(3) | 1940(190) | 1790(180) | 18.5(14) | 20.0(20) | 5.25(10) | -0.48(15) |
| | 32(3) | 1850(180) | 1700(170) | | | 5.45(10) | -0.68(11) |
| | 32(3) | 1740(170) | 1590(160) | | | 5.69(10) | -0.92(14) |
| | 31(3) | 1660(170) | 1510(150) | | | 5.92(10) | -1.15(14) |
| | 31(3) | 1560(160) | 1420(140) | | | 6.15(10) | -1.38(14) |
| | | | | | | | |
| 2 | 67(7) | 2070(210) | 1950(200) | 9.8(7) | 20.0(20) | 5.30(0) | -0.81(11) |
| | 68(7) | 2320(230) | 2200(220) | | | 4.84(0) | -0.34(11) |
| | 67(7) | 1870(190) | 1760(180) | | | 6.21(0) | -1.72(11) |
| | 67(7) | 2010(200) | 1900(190) | | | 5.40(0) | -0.91(11) |
| | 68(7) | 2190(220) | 2070(210) | | | 5.06(0) | -0.56(11) |
| | 67(7) | 2020(200) | 1900(190) | | | 5.35(0) | -0.86(11) |
| | 67(7) | 1900(190) | 1790(180) | | | 6.09(0) | -1.59(11) |
| | 67(7) | 1780(180) | 1670(170) | | | 6.62(0) | -2.13(11) |
| | 66(7) | 1610(160) | 1500(150) | | | 7.01(0) | -2.52(11) |

**Table 2. Arrhenius fitting parameters for the electrical conductivity of davemaoite.**

| Region | Pressure (GPa) | Run# | $T$ range (K) | $\Delta H$ (eV) | $\log\sigma_0$ (S/m) | $R^2$ |
|---|---|---|---|---|---|---|
| I | 31–32 | 2 | 1420–1790 | 1.23(3) | 2.98 | 0.999 |
| | 43–48 | 1 | 1040–1440 | 0.95(2) | 1.59 | 0.999 |
| | 50–57 | 1 | 1080–1550 | 0.95(2) | 1.05 | 0.998 |
| | 66–68 | 2 | 1500–1670 | 1.14(45) | 1.31 | 1.000 |
| | 83–89 | 1 | 1210–1740 | 1.15(4) | 1.23 | 0.992 |
| II | 43–48 | 1 | 1490–1620 | 1.45(19) | 3.36 | 0.967 |
| | 50–57 | 1 | 1620–1770 | 2.05(19) | 4.55 | 0.982 |
| | 66–68 | 2 | 1760–1900 | 4.07(22) | 9.91 | 0.994 |
| | 83–89 | 1 | 1800–1880 | 2.32(65) | 4.53 | 0.927 |
| III | 43–48 | 1 | 1680–1890 | 1.04(15) | 2.08 | 0.960 |
| | 50–57 | 1 | 1850–2010 | 1.05(10) | 1.70 | 0.991 |
| | 66–68 | 2 | 1950–2200 | 1.60(4) | 3.33 | 0.999 |

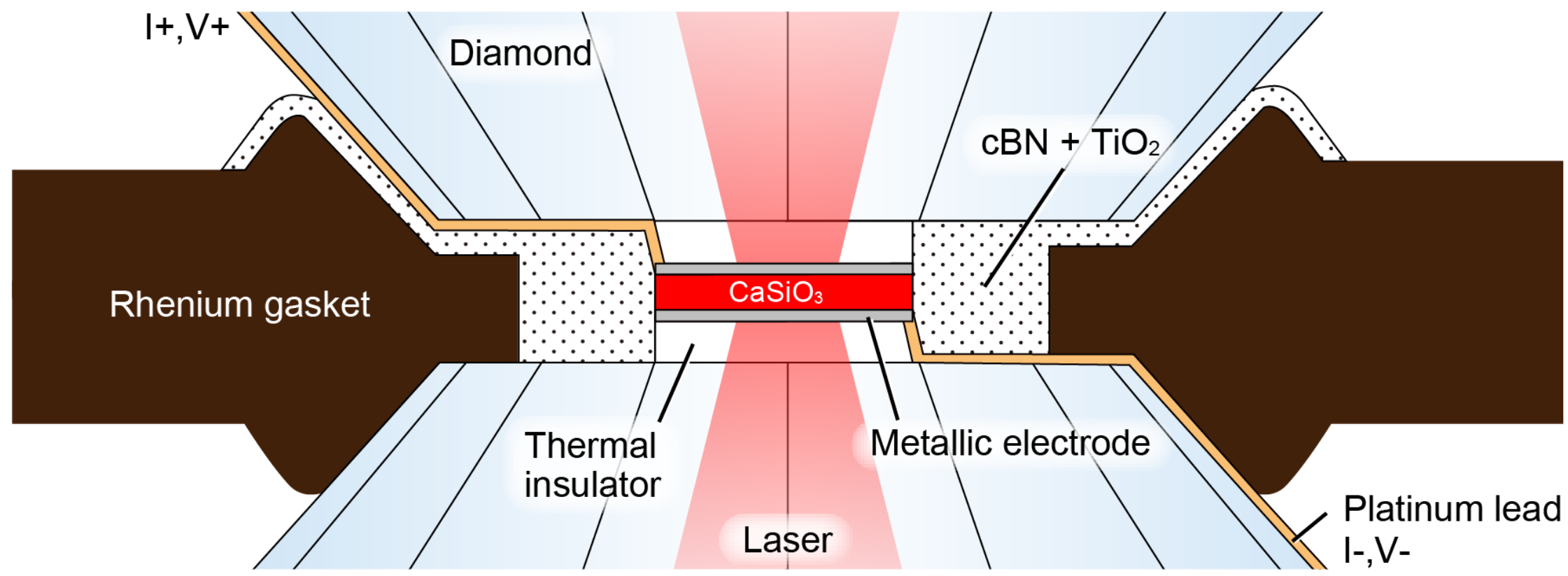

**Fig. 1.** Schematic illustration of the cell assembly for electrical conductivity measurements. Pt leads were connected to copper wires and an impedance analyzer.

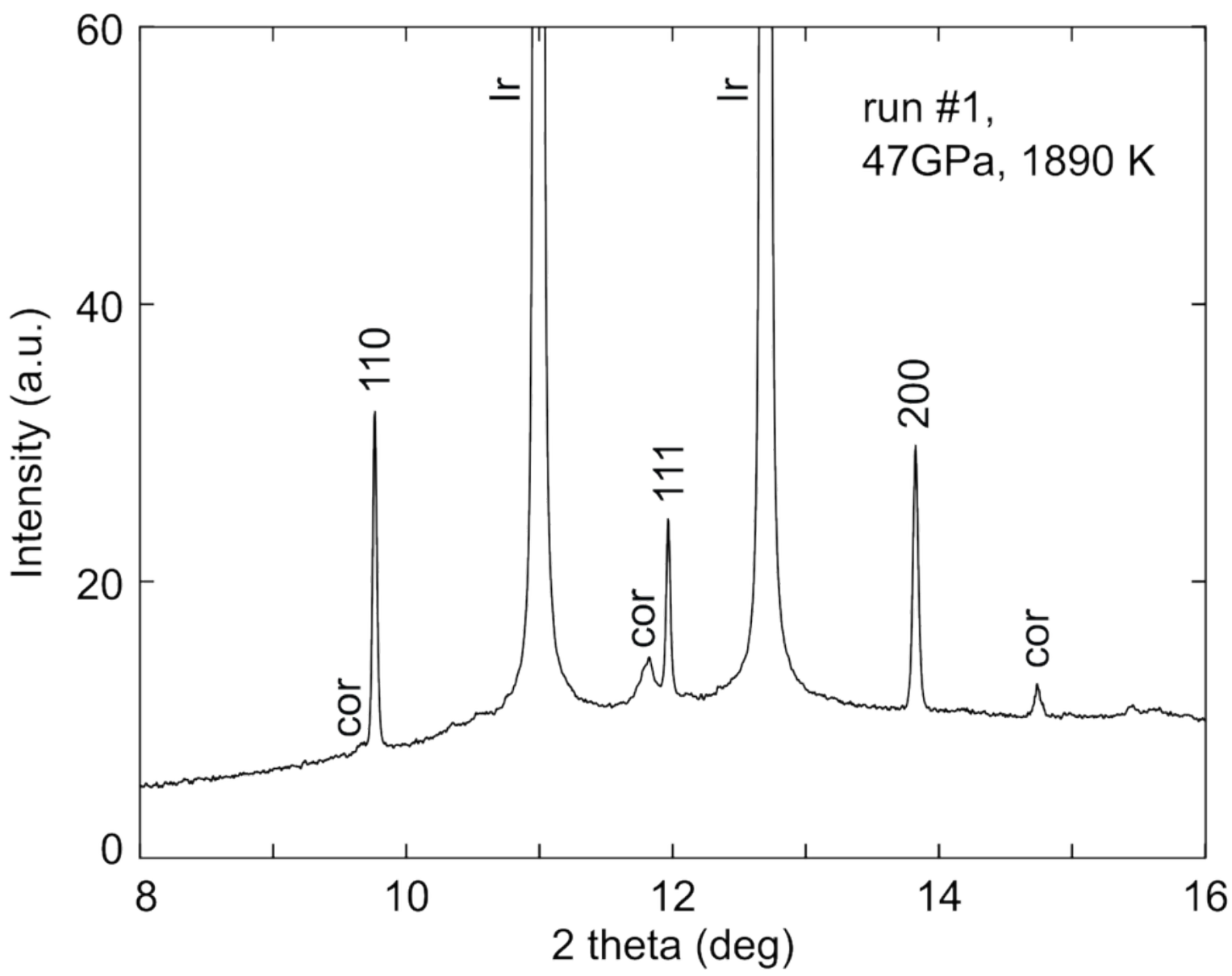

**Fig. 2.** Typical XRD pattern collected at 47 GPa and 1890 K in run #1. Numbers indicate Miller indices of the $CaSiO_3$ perovskite phase.

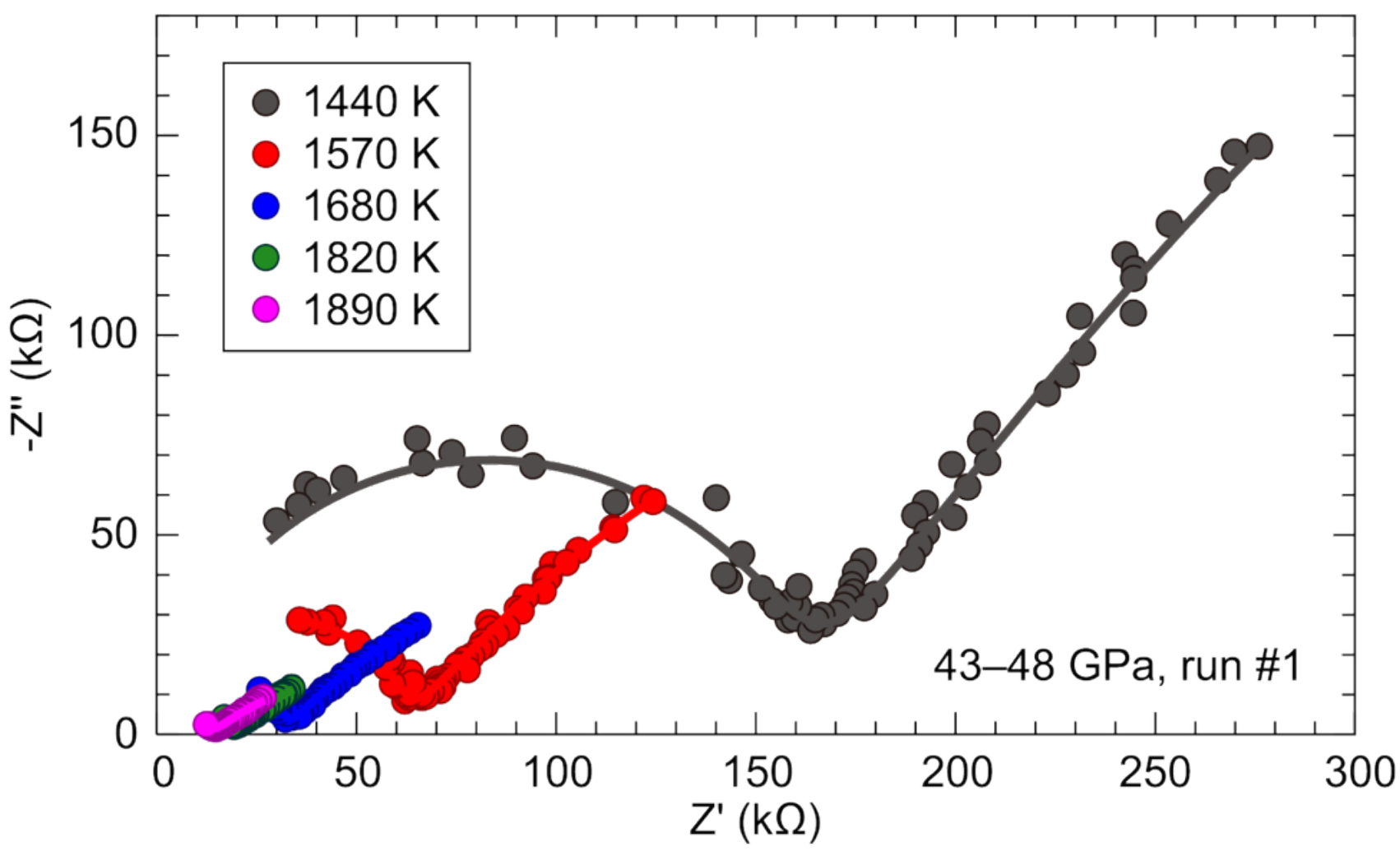

**Fig. 3.** High-pressure and high-temperature impedance spectra collected in the frequency range of $10^1$–$10^6$ Hz at 43–48 GPa and 1440–1890 K in run #1.

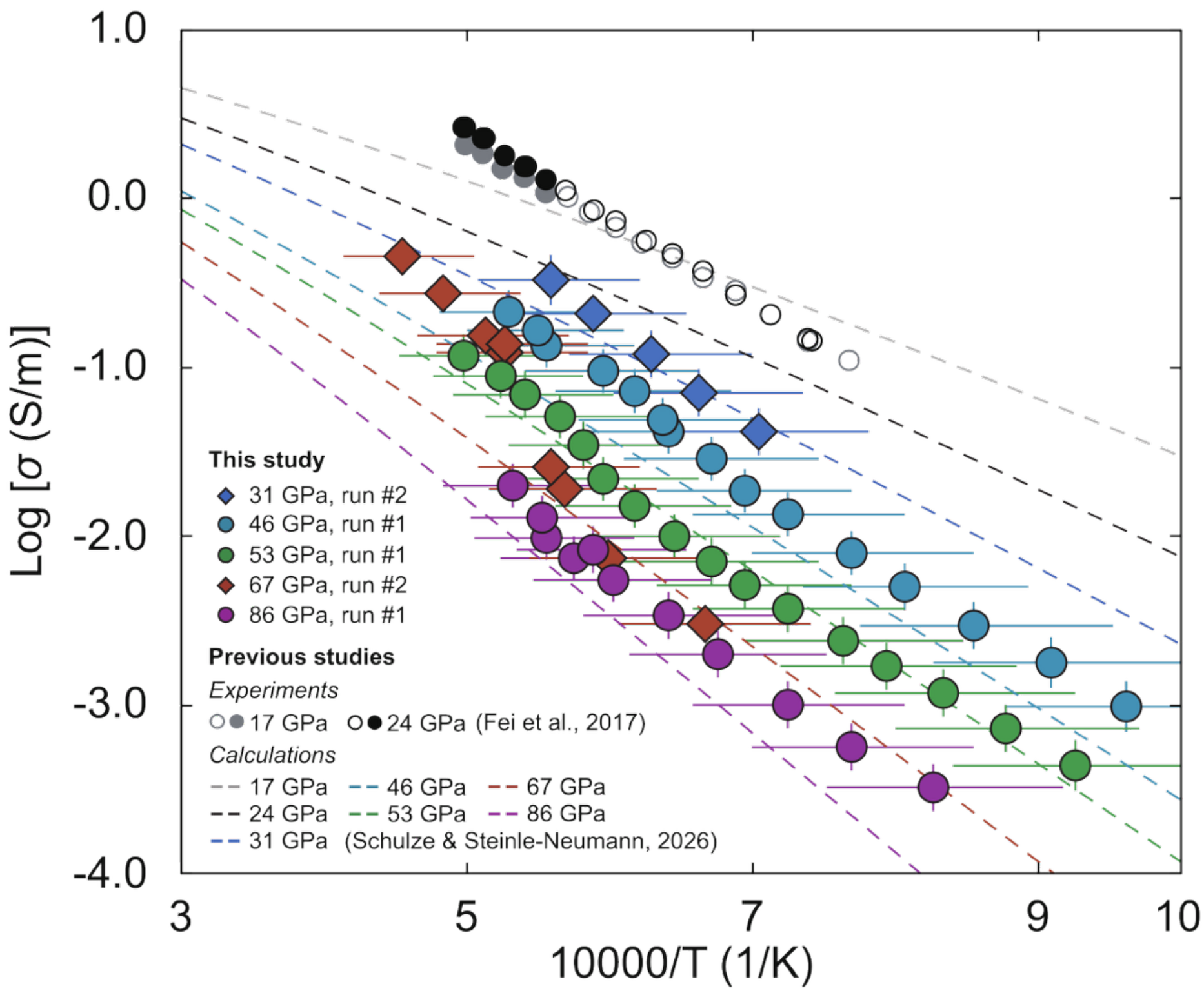

**Fig. 4.** Arrhenius plot of the electrical conductivity of $CaSiO_3$ perovskite at high pressure and temperature. Circles show the results from run #1 at approximately 46 GPa (cyan), 53 GPa (green), and 86 GPa (magenta). Diamonds show the results from run #2 at approximately 31 GPa (blue) and 67 GPa (brown). Gray and black circles show previous experimental measurements at 17 and 24 GPa, respectively (Fei et al., 2017). Open and filled symbols denote the datasets interpreted as protonic and ionic conduction, respectively, in that study. Dashed lines show the calculated electrical conductivities associated with oxygen-vacancy diffusion at 17, 24, 31, 46, 53, 67, and 86 GPa (Schulze and Steinle-Neumann, 2026).

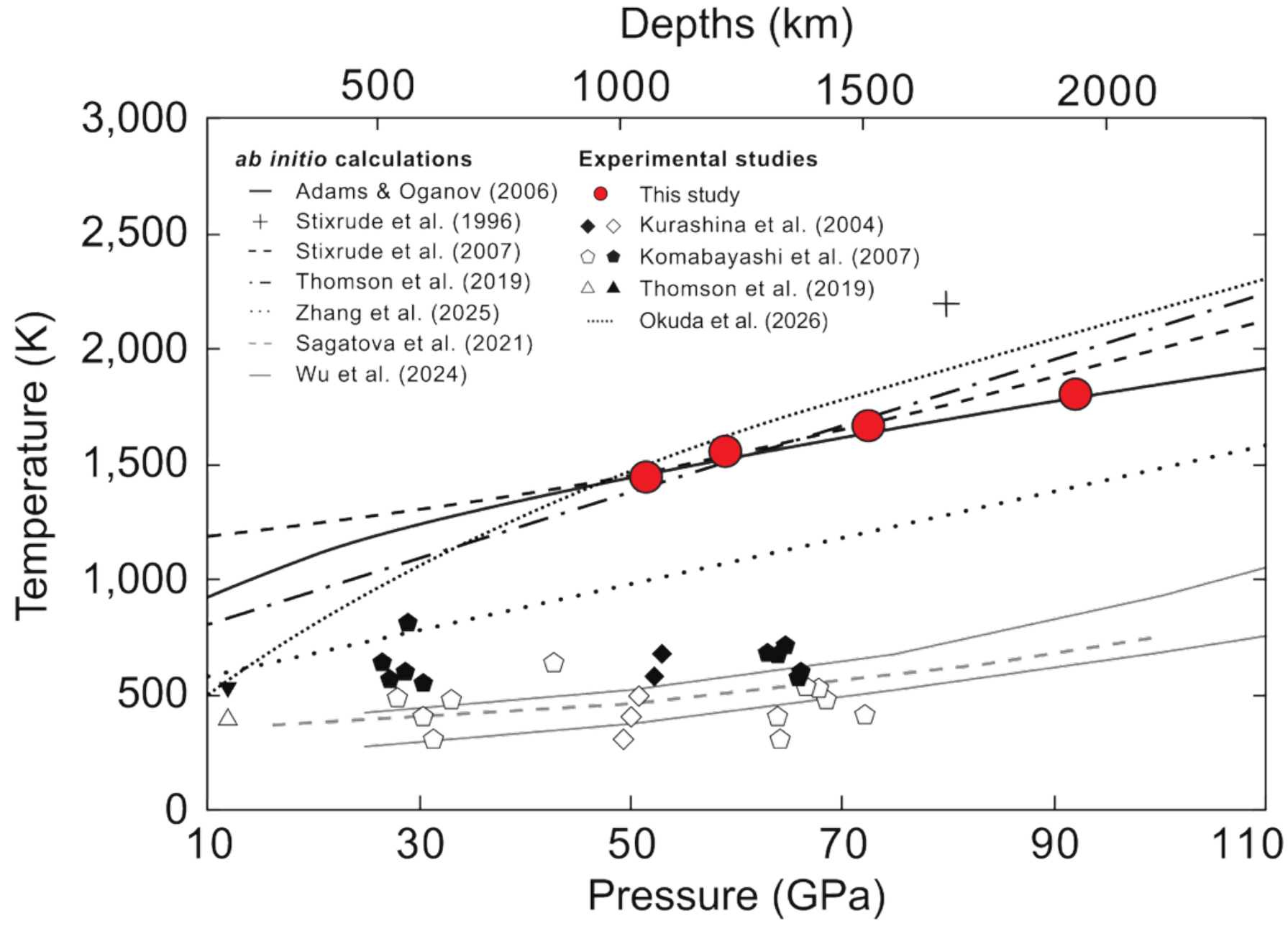

**Fig. 5.** Pressure–temperature conditions of the tetragonal-to-cubic transition in $CaSiO_3$ perovskite. Red open circles indicate the *P*–*T* conditions at which a distinct change in the temperature dependence of electrical conductivity was observed in this study. Diamonds, pentagons, and triangles show previous *in situ* XRD results of Kurashina et al. (2004), Komabayashi et al. (2007), and Thomson et al. (2019), respectively. The finely dotted curve denotes the transition determined by the *in situ* XRD study of Okuda et al. (2026a). Open and filled symbols indicate the tetragonal and cubic structures, respectively. Solid, broken, dash-dotted, dotted, dashed, and thin solid curves, and a plus symbol indicate the tetragonal-to-cubic transition predicted by Adams and Oganov (2006), Stixrude et al. (2007), Thomson et al. (2019), Zhang et al. (2025), Sagatova et al. (2021), Wu et al. (2024), and Stixrude et al. (1996), respectively.

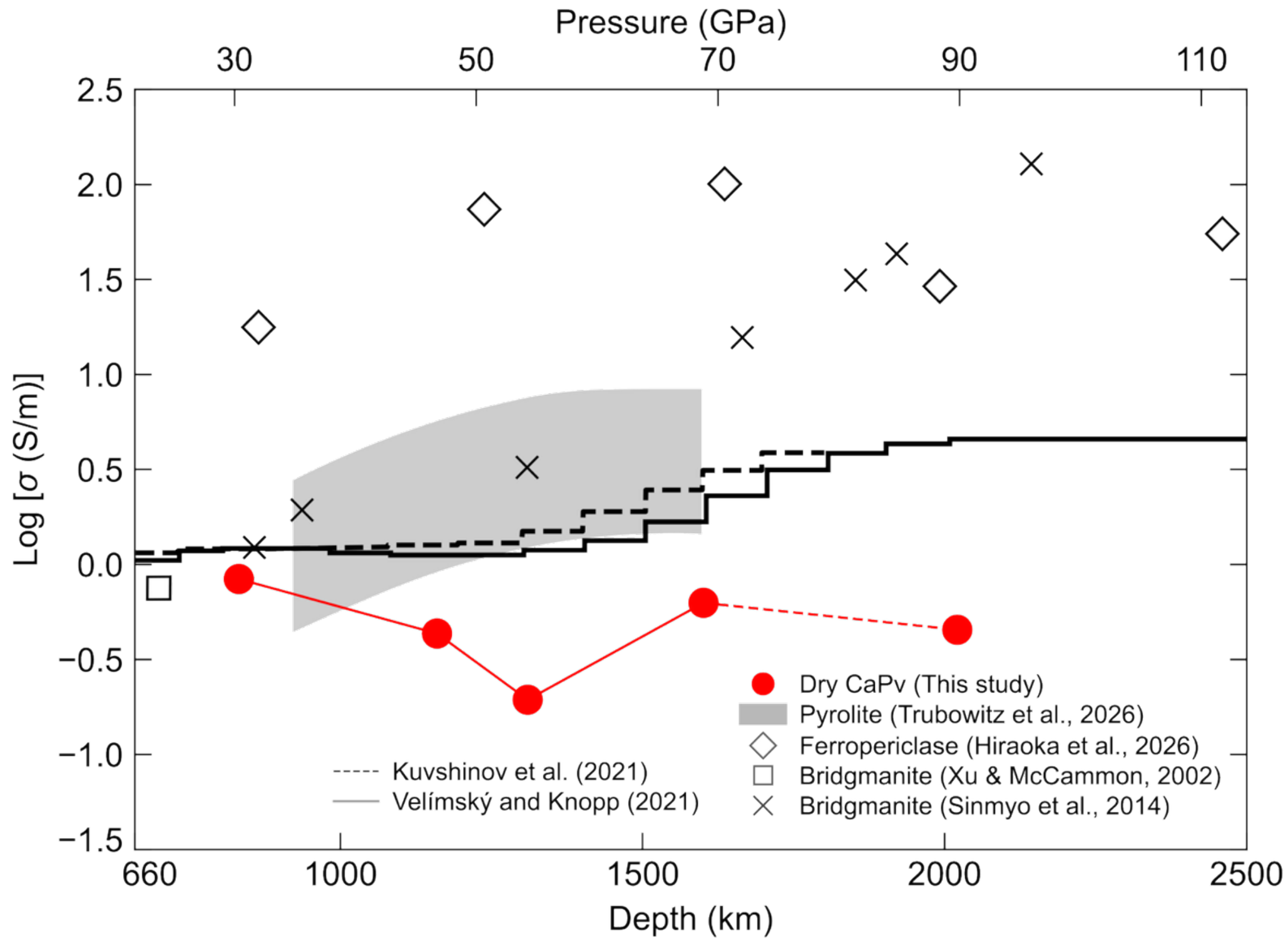

**Fig. 6.** Electrical conductivity of dry davemaoite calculated along the normal lower-mantle geotherm (Katsura, 2022). Red circles show representative conductivity values calculated from the Arrhenius relations obtained in this study, with depth and pressure shown on the lower and upper axes, respectively. The dashed red extension to 89 GPa shows the conductivity obtained by extrapolating the Region II Arrhenius relation to the corresponding geotherm temperature. Previous estimates of lower-mantle conductivity and conductivities of major lower-mantle materials are shown for comparison, including geophysical conductivity models (Kuvshinov et al., 2021; Velímský and Knopp, 2021), pyrolite (Ohta et al., 2010; Trubowitz et al., 2026), ferropericlase (Hiraoka et al., 2026), and (Fe,Al)-bearing bridgmanite (Sinmyo et al., 2014; Xu and McCammon, 2002).

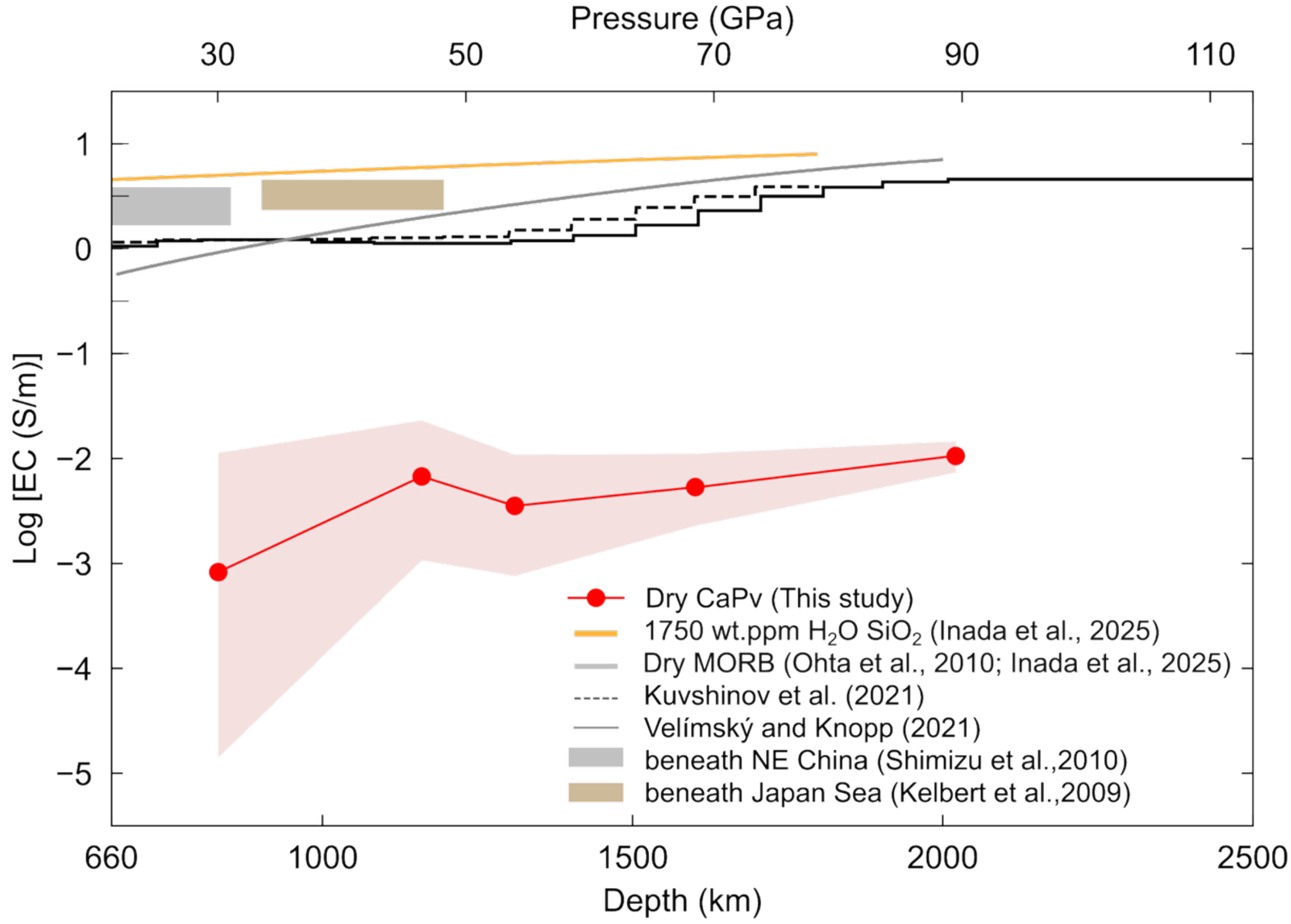

**Fig. 7.** Electrical conductivity of dry davemaoite along a cold slab geotherm compared with previous experimental results and geophysical observations. Filled red circles show the EC of dry davemaoite calculated at representative temperatures along the cold slab geotherm, and the red shaded region indicates the range of calculated EC corresponding to the cold slab geotherm (Ohtani, 2020). Previous results for hydrous $SiO_2$ containing 1750 wt. ppm $H_2O$ (Inada et al., 2025) and dry MORB (Ohta et al., 2010) are shown for comparison. Shaded boxes indicate high-conductivity anomalies observed beneath northeastern China (Shimizu et al., 2010) and the Japan Sea (Kelbert et al., 2009).

# Supplementary Material

## Low electrical conductivity of dry $CaSiO_3$ perovskite under lower mantle conditions

Yoshiyuki Okuda[1,2,*], Bin Chen[1], Juliana Peckenpaugh[1,3], Hirokazu Kadobayashi[4]

[1] *Hawai'i Institute of Geophysics and Planetology, University of Hawai'i at Manoa, Honolulu, Hawai'i 96822, USA*

[2] *Department of Earth and Planetary Sciences, Institute of Science Tokyo, Meguro, Tokyo 152-8551, Japan*

[3] *Department of Earth Sciences, University of Hawai'i at Manoa, Honolulu, Hawai'i 96822, USA*

[4] *SPring-8, Japan Synchrotron Radiation Research Institute, Sayo, Hyogo 679-5198, Japan*

*Corresponding author

*E-mail:* Yoshiyuki Okuda (yokuda@hawaii.edu)

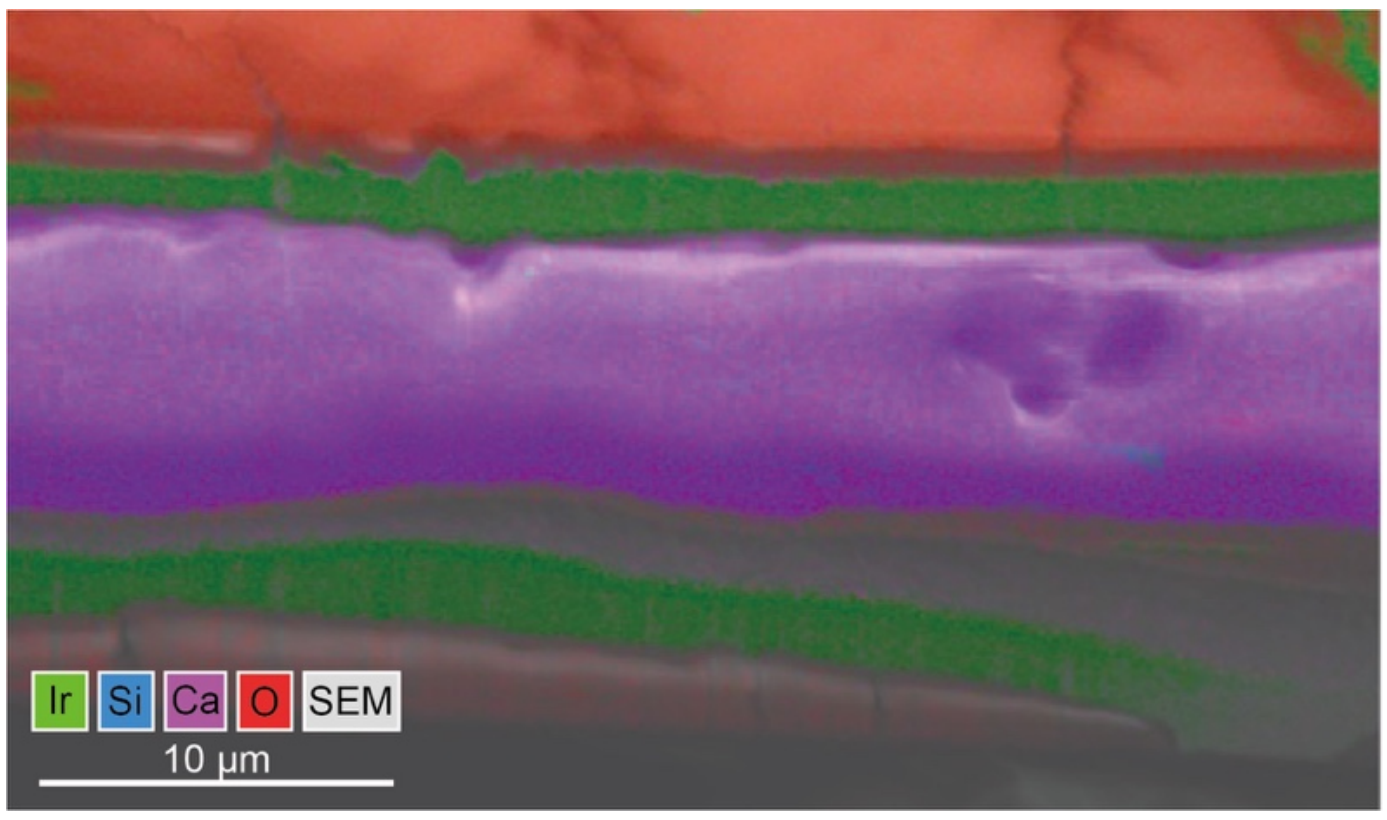

**Fig. S1.** Elemental map of a cross section of the sample recovered from run #2.

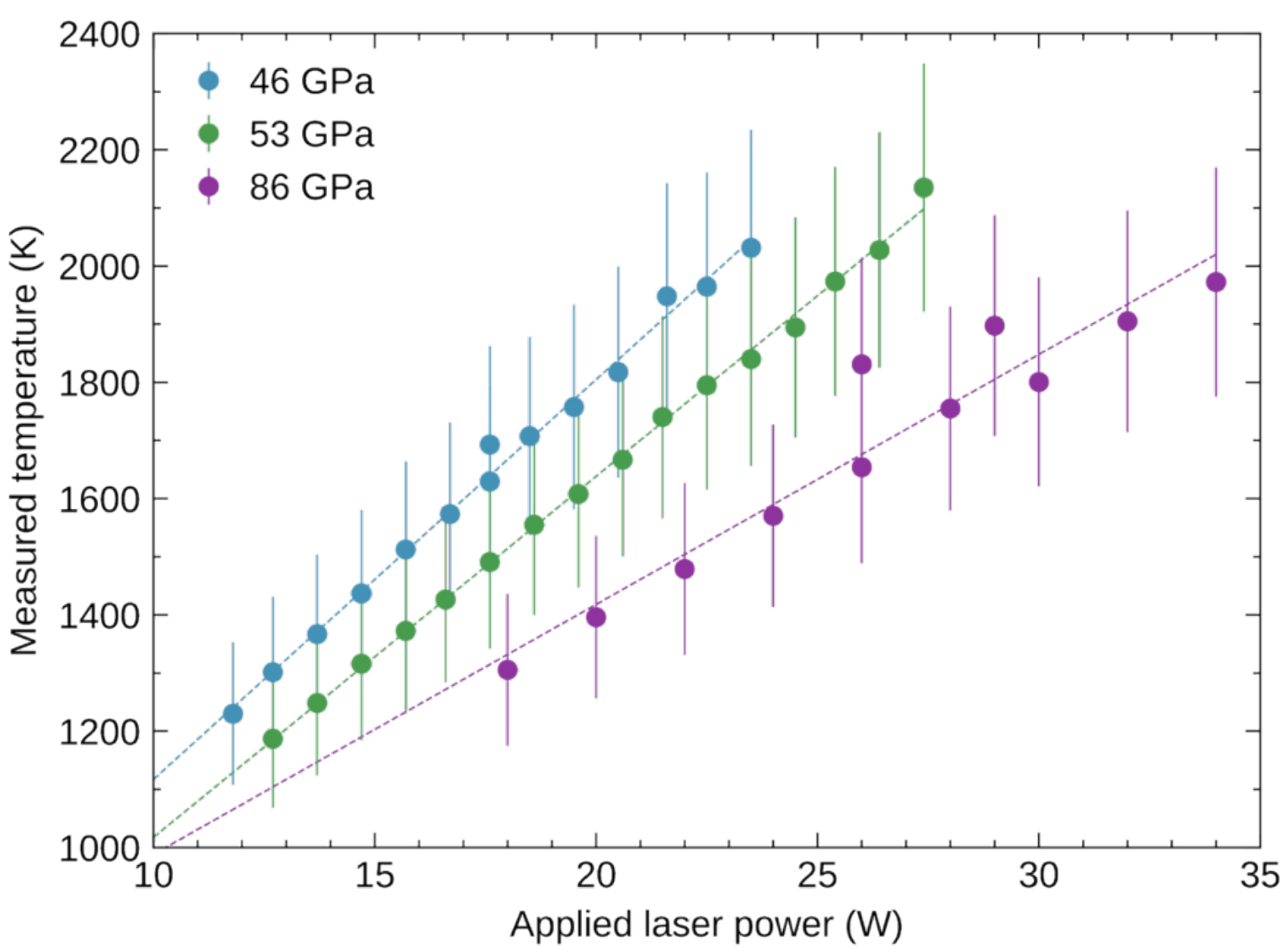

**Fig. S2.** Relationship between applied laser power and measured temperature in run #1. Dashed lines show linear least-squares fits to the measured temperatures. The fitted relationships were used to estimate temperatures at lower laser powers where thermal emission was insufficient for direct pyrometric measurement.

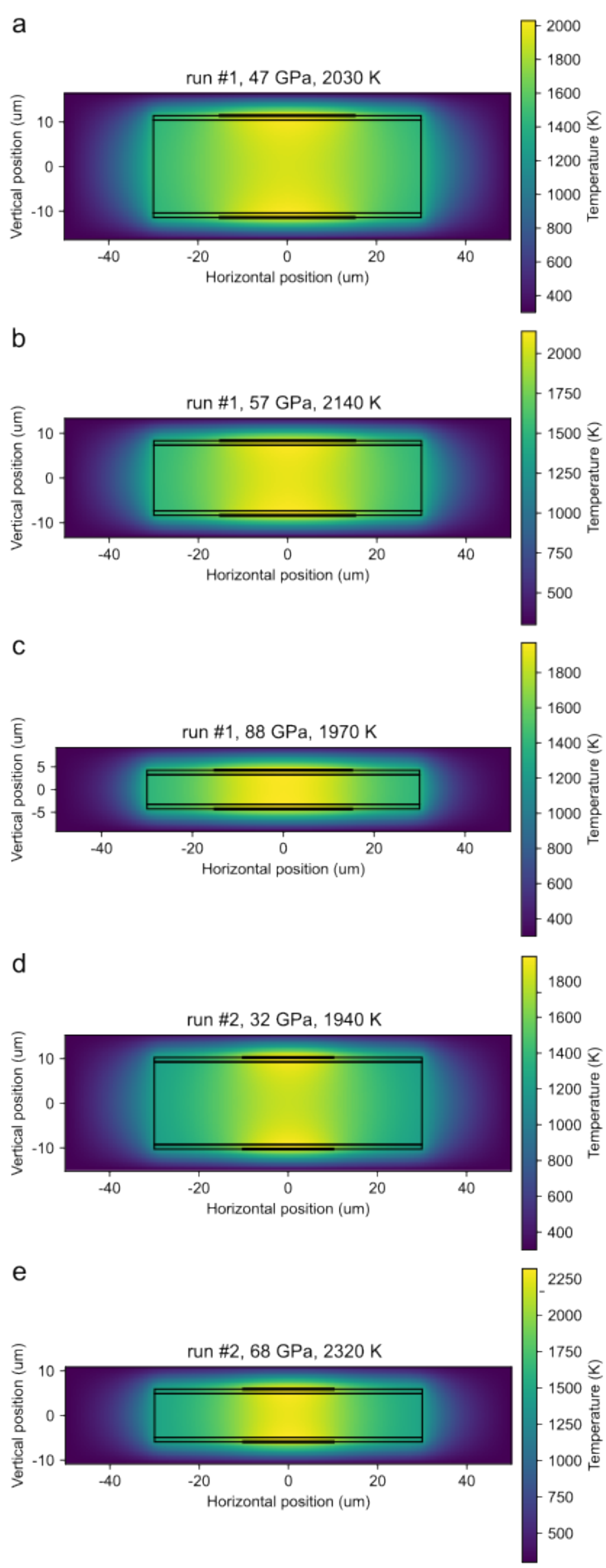

**Fig. S3. Calculated temperature distributions under the experimental conditions.** (a–c) Run #1 at 47, 57, and 88 GPa, respectively; (d–e) run #2 at 32 and 68 GPa, respectively. Each panel corresponds to the highest measured surface temperature at each pressure condition. Black line segments indicate the laser-irradiated width.

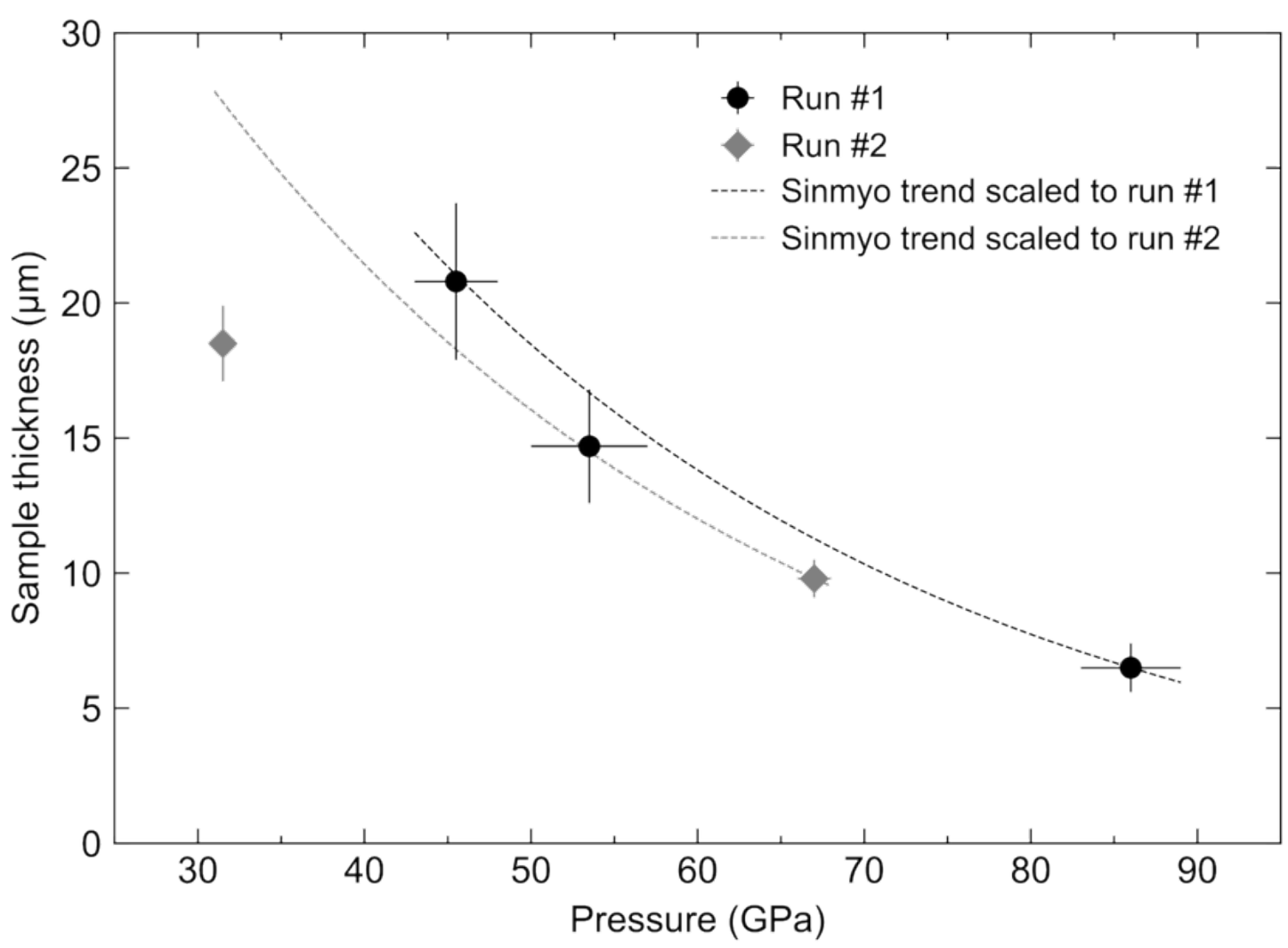

**Fig. S4.** Pressure dependence of sample thickness used in this study. Symbols show the thicknesses adopted for run #1 (black circles) and run #2 (gray diamonds). For comparison, dashed curves show the pressure-dependent trend reported by Sinmyo et al. (2014), scaled to the highest-pressure thickness of each run.

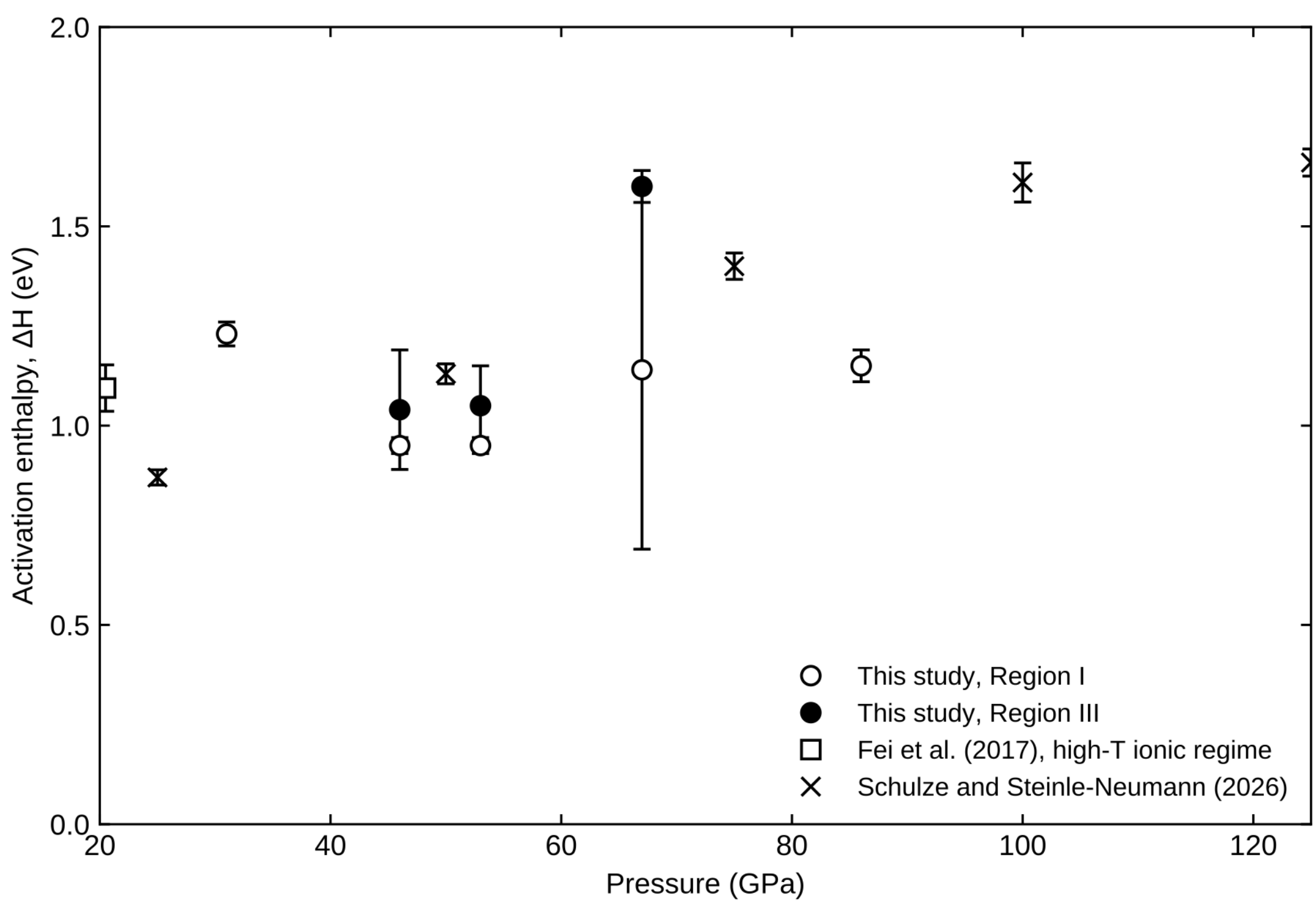

**Fig. S5.** Activation enthalpy, $\Delta H$, as a function of pressure for davemaoite. Open and filled circles show the values for Regions I and III, respectively, determined in this study. The open square shows $\Delta H$ for the high-temperature ionic-conduction regime of Fei et al. (2017), calculated from their reported $\Delta E$ and $\Delta V$ using $\Delta H = \Delta E + P\Delta V$ at a representative pressure of 20.5 GPa. Crosses show calculated $\Delta H$ values for extrinsic oxygen-vacancy diffusion in cubic davemaoite (Schulze and Steinle-Neumann, 2026).

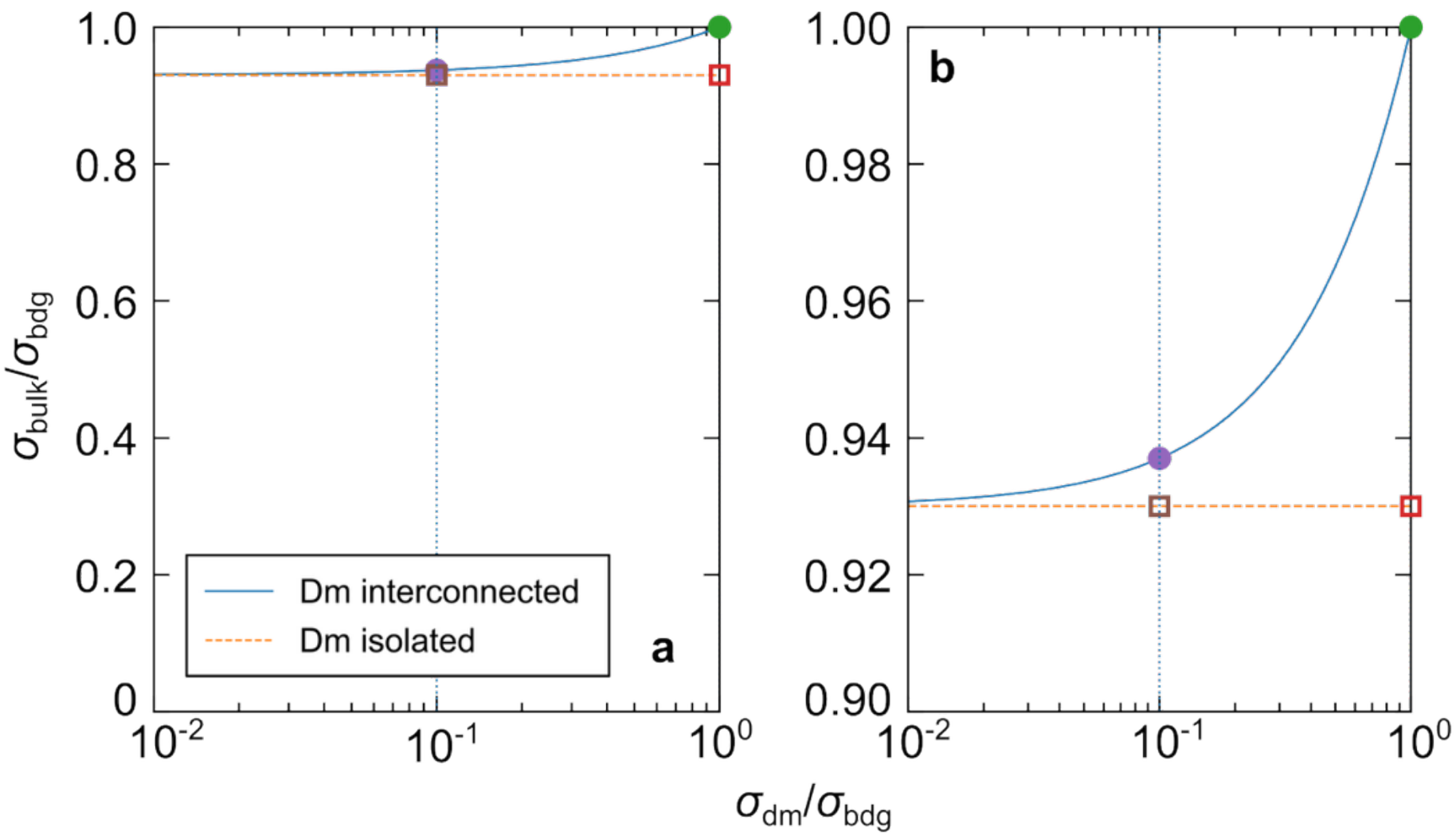

**Fig. S6.** Calculated bulk electrical conductivity of a bridgmanite-davemaoite aggregate as a function of the conductivity ratio between davemaoite and bridgmanite. The calculations assume a two-phase aggregate containing 93 vol% bridgmanite (bdg) and 7 vol% davemaoite (dm), corresponding to an upper-bound pyrolitic abundance of davemaoite (Hao et al., 2026; Irifune et al., 2010). (a) Full-range view. (b) Enlarged view for $\sigma_{bulk}/\sigma_{bdg} = 0.9–1.0$. Two end-member connectivity scenarios are considered. In the interconnected model, both bridgmanite and davemaoite are assumed to form continuous conducting pathways, and the bulk conductivity is calculated using the parallel mixing relation $\sigma_{bulk} = 0.93\sigma_{bdg} + 0.07\sigma_{dm}$. In the electrically isolated (non-contributing) end-member, davemaoite is assumed not to contribute to macroscopic electrical conduction, such that conduction is carried only by the 93 vol% bridgmanite fraction, yielding $\sigma_{bulk} = 0.93\sigma_{bdg}$. Filled circles and open squares indicate representative conditions relevant to this study: $\sigma_{dm}/\sigma_{bdg} \approx 1$ at 31 GPa, where the conductivity of davemaoite is comparable to that of bridgmanite, and $\sigma_{dm}/\sigma_{bdg} \approx 0.1$ as a representative higher-pressure case, where davemaoite is substantially less conductive than bridgmanite. In both end-member scenarios, the calculated bulk conductivity remains close to that of bridgmanite, indicating that dry davemaoite makes a negligible contribution to the electrical conductivity of a pyrolitic lower mantle, particularly considering that its actual abundance in the ambient lower mantle may be lower than 7 vol% (Hao et al., 2026).

**Supplementary Text S1. Calculation of theoretical ionic conductivity of davemaoite**

The theoretical ionic conductivity of davemaoite was calculated from the oxygen-vacancy diffusion model of Schulze and Steinle-Neumann (2026). Their simulations indicate that the oxygen-vacancy diffusion coefficient, $D_\mathrm{v}$, follows an Arrhenius-type relation,

$$D_\mathrm{v}(P,T) = D_\mathrm{v}^0 \exp\left[-\frac{\Delta H_\mathrm{a}(P)}{k_\mathrm{B}T}\right], \qquad \text{(S1)}$$

where $P$ is pressure, $T$ is temperature, $D_\mathrm{v}^0$ is the pre-exponential factor, $\Delta H_\mathrm{a}$ is the pressure-dependent activation enthalpy, and $k_\mathrm{B}$ is the Boltzmann constant. The pressure-independent pre-exponential factor is

$$\log_{10}[D_\mathrm{v}^0\ (\mathrm{m^2\ s^{-1}})] = -6.53, \qquad \text{(S2)}$$

and the activation enthalpy is parameterized as

$$\Delta H_\mathrm{a}(P) = 0.779\ln(P + 25.594) - 2.155, \qquad \text{(S3)}$$

where $P$ is expressed in GPa and $\Delta H_\mathrm{a}$ is expressed in eV.

The ionic conductivity was calculated from $D_\mathrm{v}$ using the Nernst-Einstein relation,

$$\sigma_\mathrm{ion} = \frac{q^2 C_\mathrm{v} D_\mathrm{v}}{k_\mathrm{B} T R_\mathrm{H}}, \qquad \text{(S4)}$$

where $q$ is the charge of the mobile defect, $C_\mathrm{v}$ is the oxygen-vacancy number density, and $R_\mathrm{H}$ is the Haven ratio. Oxygen vacancies were treated as doubly charged defects, such that $q = 2e$, where $e = 1.602176634 \times 10^{-19}$ C. Following Schulze and Steinle-Neumann (2026), we assumed $R_\mathrm{H} = 1$, and an oxygen-vacancy concentration of $C_\mathrm{v} = 10^{20}$ $\mathrm{cm^{-3}}$. Schulze and Steinle-Neumann (2026) examined vacancy concentrations between $10^{19}$ and $10^{21}$ $\mathrm{cm^{-3}}$ and identified approximately $10^{20}$ $\mathrm{cm^{-3}}$ as a representative order-of-magnitude value based on plausible monovalent impurity concentrations in davemaoite. The theoretical conductivity curves were calculated at pressures of 17, 24, 31, 46, 53, 67, and 86 GPa. The curves at 31, 46, 53, 67, and 86 GPa correspond to the experimental pressure conditions of this study, whereas the 17 and 24 GPa curves were calculated for comparison with the electrical conductivity measurements of previous studies (Fei et al., 2017). For each pressure, the temperature-dependent diffusion coefficient was first calculated using Eqs. (S1) to (S3), and the resulting diffusion coefficient was then converted to ionic conductivity using Eq. (S4). The calculated conductivity was plotted as $\log_{10}[\sigma_\mathrm{ion}\ (\mathrm{S\ m^{-1}})]$ against 10000/$T$.

The calculated activation enthalpies are 0.768, 0.886, 0.989, 1.172, 1.245, 1.372, and 1.518 eV at 17, 24, 31, 46, 53, 67, and 86 GPa, respectively. These theoretical curves were calculated independently of the experimental conductivity data and were not fitted to either the present measurements or the data of Fei et al. (2017). The absolute magnitude of the calculated conductivity depends linearly on the assumed oxygen-vacancy concentration. Consequently, increasing or decreasing $C_v$ by one order of magnitude shifts $\log_{10}\sigma_{\mathrm{ion}}$ upward or downward by one unit, respectively, without changing the slope of the Arrhenius relation. The calculated values should therefore be regarded as model predictions for the specified assumptions of $C_v = 10^{20}$ cm$^{-3}$ and $R_H = 1$, rather than as concentration-independent theoretical conductivities. The original study also notes that uncertainties in the vacancy concentration and vacancy interactions affect the absolute conductivity estimates (Schulze and Steinle-Neumann, 2026).

**Supplementary References**